\documentclass[12pt]{article}
\usepackage{graphicx}
\def\={~=~}
\def\+{~+~}
\def\-{~-~}

\def\tilde{\widetilde}
\def\beq{\begin{equation}}
\def\eeq{\end{equation}}
\def\beqn{\begin{eqnarray}}
\def\eeqn{\end{eqnarray}}
\newcommand{\calle}[1]{(\ref{#1})}

\def\At{{\tilde A_1}}
\def\Atd{{\tilde A_1^\dagger}}
\def\Chat{{\tilde C}}
\def\hatC{{\tilde C}}
\def\bareta{{\tilde\eta}}

\def\bub{{\bf U}}
\def\bab{{\bf A}}
\def\sech{{\rm sech}}
\def\tanh{{\rm tanh}}

\newcommand{\bye}{\end{document}}

\newcommand{\beql}[1]{\beq}

\newcommand{\beqnl}[1]{\beqn}

\def\mat#1{\left(\matrix{#1}\right)}

\def\ba{\left(\begin{array}}
\def\ea{\end{array}\right)}

\begin{document}
\rightline{HWS-2004A01}
\rightline{quant-ph/yymmddd}

\vspace{1cm}

\begin{center}
{\Large\bf 
A BPS Interpretation of Shape Invariance}
\end{center}

\vspace{5mm}
\begin{center}
Michael Faux$^\dagger$ and
Donald Spector$^{\ddagger}$\\
  Department of Physics, Eaton Hall\\
  Hobart and William Smith Colleges\\
  Geneva, NY 14456, USA
\end{center}

\vspace{2cm}

\begin{abstract}
\noindent
We show that shape
invariance appears when a quantum mechanical model is invariant
under a centrally extended superalgebra endowed with an
additional symmetry generator,
which we dub the shift operator.  The familiar mathematical and
physical results of shape 
invariance then arise from the BPS structure associated with
this shift operator.  The shift operator 
also ensures that there is a one-to-one correspondence between
the energy levels of such a model and the energies of the
BPS-saturating states.  These findings thus provide a more comprehensive
algebraic setting for understanding shape invariance.
\end{abstract}
\centerline{January 2004}

\vfill
\hrule
\noindent
$\dagger$ e-mail address: faux@hws.edu\\
$\ddagger$ e-mail address: spector@hws.edu\\

\newpage

\section{Introduction}

Shape invariance \cite{shapeinv} provides perhaps 
the most illuminating approach
to exact solubility in quantum mechanics.  Building on the properties
of supersymmetric quantum mechanics \cite{WittenSQM}, 
shape invariance offers an elegant
and concise algorithm for generating the stationary states and energy
eigenvalues in exactly solvable models.  The understanding
of this property, however, has in many ways remained incomplete.  Why does it appear
when it does? Why
are some models shape invariant and others not?  

In this paper, we identify the deeper structure that produces
shape invariance.  Shape invariance arises when a quantum mechanical
model is invariant under both a supersymmetry algebra with a central
charge and an additional symmetry operator, analogous to a LaPlace-Runge-Lenz
vector \cite{rungelenz}.  The results of shape invariance can then be understood
as arising from the BPS-like \cite{bps} phenomena associated with this additional
operator,
with the added feature that every state in the theory is degenerate
with, and easily obtainable from, one of the BPS states of the
model.  

This larger algebra suggests 
that shape invariance  may well have a much broader role to
play in physics, since
centrally extended superalgebras have come to be of great
importance in field theory and string theory.  In addition,
the appearance of BPS bounds
and equations indicates the presence of an
underlying topological
structure.  We thus expect our work to provide a
framework for identifying the appearance of shape
invariance and its associated properties in other settings of
significance.

\section{Shape Invariance Reviewed}
\label{sec:2}

Consider non-relativistic quantum mechanics in one spatial
dimension.  When the potential energy is adjusted so that 
the ground state energy is zero, the Hamiltonian can be 
written in a factorized form
\beq
\label{H1AA}
H_1(g) \= A^\dagger(g)A(g)~~,
\eeq
where $g$ denotes the real parameter(s) that determine the
potential, and $A(g)$ is a first-order
differential operator.  This Hamiltonian is positive semi-definite, and
its ground state wavefunction is 
the state annihilated by $A(g)$.

Reversing the order of $A$ and $A^\dagger$ in \calle{H1AA}
produces an affiliated ``partner'' Hamiltonian
\beq
\label{H2AA}
H_2 \= A(g) A^\dagger(g)~~.
\eeq
The only difference between the spectra of $H_1$ and $H_2$ is
that $H_1$ has a zero-energy state and $H_2$ in general does
not; otherwise, their spectra are identical.  To
see that the positive energy spectra of these two Hamiltonians
are degenerate, observe that $H_2 A = A H_1$.  Consequently,
if $H_1 \psi = E\psi$ for a wavefunction $\psi$ not annihilated by $A$,
then $A\psi$ is an eigenstate of $H_2$ satisfying
$H_2 (A\psi) = E(A\psi)$.  Likewise, $A^\dagger$
maps eigenstates of $H_2$ to degenerate eigenstates of $H_1$.

Supersymmetry provides a natural context for understanding 
the relationships between the states of $H_1$ and those of $H_2$.
If one combines these two operators into
\beq
H\= \left(\matrix{H_1 & 0\cr 0 & H_2}\right)~~,
\eeq
this matrix Hamiltonian can be obtained from the
anticommutator $H=\{Q,Q^\dagger\}$, where $Q$
and $Q^\dagger$ are supercharges, given by
\beq
Q=\left(\matrix{0 & 0\cr A & 0}\right)
\quad , \qquad
Q^\dagger=\left(\matrix{0 & A^\dagger\cr 0 & 0}\right)~~.
\eeq
Both $Q$ and $Q^\dagger$ commute with $H$. The operator 
$\Gamma = \sigma_3$
has eigenvalues $\pm 1$ that distinguish the $H_1$ and $H_2$ sectors.  Since
$\{Q,\Gamma\}=\{Q^\dagger,\Gamma\}=0$,  the supercharges $Q$ and
$Q^\dagger$ map
states from one $\Gamma$-sector into the degenerate states of the other 
$\Gamma$-sector.
The operator $\Gamma$ thus plays a role in supersymmetric quantum
mechanics analogous to the role played by the operator $(-1)^F$ in
supersymmetric field theories \cite{Windex}.

Shape invariance is a property that arises when there is an additional
relationship between the partner Hamiltonians $H_1$ and $H_2$.  Suppose
that these Hamiltonians are linked by the condition
\beq
\label{ShapeCondition}
A(g_1)A^\dagger(g_1) = A^\dagger(g_2)A(g_2)+c(g_2)~~~,
\eeq
where the real parameters $g_1$ and $g_2$ are related by a mapping
$f:g_1\rightarrow g_2$, and $c(g)$ is a $c$-number that
depends on the parameter(s) of the Hamiltonian. 
When this condition holds,
the Hamiltonian $H_1$ is said to be
shape invariant.

One can readily determine the states and energy levels of a shape invariant
Hamiltonian.  Denote the
energy levels of $H_1$ by $E_n$,
and those of $H_2$ by
$\tilde E_n$, where $n=1,2,3,\ldots~$.  (The label $n$ gives the levels in order of
increasing energy, with $n=1$ corresponding to the ground state.)  
Then \calle{ShapeCondition} implies
that $\tilde E_n(g_1) = E_n(g_2)+c(g_2)$, while supersymmetry
implies $E_{n+1}(g_1) = \tilde E_n(g_1)$.  
Supersymmetry also provides a
map between the level $n$ wavefunction of $H_1$ and the level
$n-1$ wavefunction of $H_2$.  Altogether, these results enable one
to solve for the spectrum of a shape invariant Hamiltonian.

Thus, for example, the ground state of $H_1$
is the function $\psi_1(x;g_1)$ annihilated by $A(g_1)$.
Because of \calle{ShapeCondition},
the ground state of $H_2$ is thus given by $\psi_1(x;g_2)$,
which is annihilated by $A(g_2)$ and has energy $c(g_2)$.  This
implies in turn that the first excited 
state of $H_1$ is $A^\dagger(g_1)\psi_1(x;g_2)$
and that this state also has energy $c(g_2)$.

The relationship \calle{ShapeCondition}
can be applied iteratively, producing a sequence of Hamiltonians
of the form
\beq
\label{Hklist}
H_k  =  A^\dagger(g_k)A(g_k)+c(g_k)+\cdots +c(g_2)~~,
\eeq
where the parameter $g_{j+1}=f(g_j)$.
Because
\beq
\label{HkAd}
A^\dagger(g_k) H_{k+1}  \=  H_k A^\dagger(g_k)~~, 
\eeq
the process described in the previous paragraph can
be iterated to obtain all the energy levels
and wavefunctions of $H_1$.
The ground state wavefunction of $H_k$ is $\psi_1(x;g_k)$,
with energy $c(g_2)+\cdots+c(g_k)$.
Applying \calle{HkAd} repeatedly, one
determines then that the
energy levels 
of the original Hamiltonian $H_1(g_1)$ are
\beq
\label{Elist}
E_n(g_1) = \sum_{j=1}^n c(g_j)~~,
\eeq
where we have defined $c(g_1)=0$; the corresponding stationary states 
are given by
\beq
\label{psilist}
\psi_n(x;g_1) = 
A^\dagger (g_1) A^\dagger(g_2)\cdots A^\dagger(g_{n-1})\psi_1(x;g_n)~~.
\eeq
Because of the mapping $f$, any parameter $g_j$ 
in the expressions in \calle{Elist} and \calle{psilist}
can be re-expressed
in terms of $g_1$.
The energy of 
the $n^{th}$ stationary state of $H_1$ is
the same as the energy of the ground state of $H_n$.

\begin{figure}
 \begin{center}
 \includegraphics[width=3in]{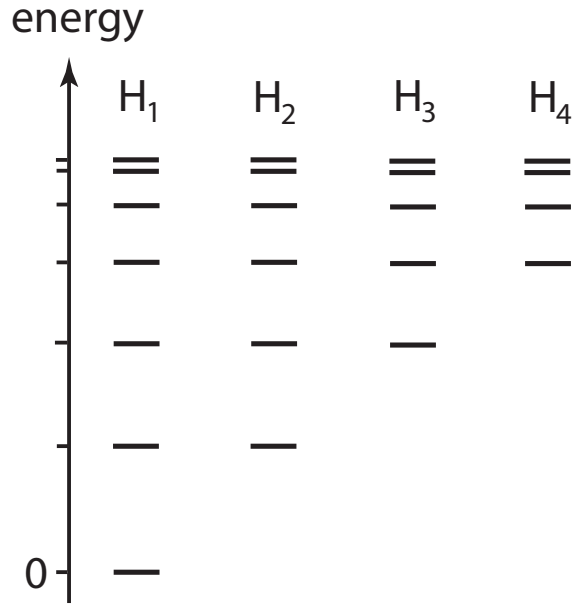}\\[.2in]
 \caption{A generic energy spectrum for a collection of four
Hamiltonians related to each other by shape invariance.  The
Hamiltonians are provided by the list  \calle{Hklist}.  Note the
degeneracies that exist across all the sectors.}
 \label{spectrum4}
 \end{center}
 \end{figure}

Figure \ref{spectrum4}  presents an illustration of the set of 
spectra that arise when we group together a set of Hamiltonians
related by shape invariance.  One notes the pervasive
degeneracies across sectors, which arise due to the shape invariance
relation \calle{ShapeCondition}.  For a concrete
realization of a shape invariant theory, we refer to the reader to the
Appendix, where we present a brief example.

Clearly, shape invariance is a useful tool for analyzing exactly
solvable quantum mechanical systems.  Why this structure should
appear in some Hamiltonians and not others, however, is not
clear, although given the intricacy and elegance of this structure, it
would seem that there ought to be an underlying principle responsible for 
its appearance.  As a first clue to understanding the origins 
of shape invariance, we 
note the following.  While
all the eigenstates of the Hamiltonians $H_k$ generated according
to \calle{Hklist} satisfy a
time-independent Schr\"odinger equation, which is second order, the
ground state of each of these sectors satisfies a simpler, first-order
equation, namely $A(g_k)\psi_1(x;g_k)=0$.  Such a scenario is familiar
from field theories in which there are BPS bounds.  In such field
theories, while the equations of
motion are generically second-order equations, the 
field configurations that saturate
BPS bounds satisfy first-order equations.  We are thus led to
consider the possibility that shape invariance is a manifestation of
BPS-saturation.  Given the close association that has been
uncovered in the field theoretic context between BPS phenomena and supersymmetry
algebras with a central charge, it would therefore seem natural to look for 
a BPS interpretation of shape invariance by considering supersymmetric quantum mechanics in
which the superalgebra includes a central charge. It is this endeavor to 
which we now turn.

\section{Supersymmetry with a Central Charge}
\label{sec:3}

Our goal in this paper is to determine the algebraic underpinnings of shape 
invariance.
As we will show, supersymmetric quantum mechanics with non-vanishing
central charge, while not sufficient to produce shape invariance, is
a key part of the framework we seek.  For now, 
we simply study quantum mechanics in which the superalgebra has
non-vanishing central charge; the connection of such centrally extended
superalgebras to shape invariance
will become apparent in the subsequent section.

To develop supersymmetric quantum mechanics with a central charge, we
first present the corresponding centrally-extended superalgebra.
For the purposes of this paper, only the case of real central charge is
relevant.  The superalgebra in this case takes the form
\beqn
\label{salg3}   
\{Q,Q^\dagger\}&=&H\nonumber \\ 
{}[H,Q] = [H,Q^\dagger]&=&0\nonumber \\
\{Q,Q\} = \{Q^\dagger,Q^\dagger\} &=&Z~~~.
\eeqn
This algebra includes the supercharges $Q$ and $Q^\dagger$, 
the real central charge $Z$,
and the Hamiltonian $H$.
When $Z=0$, the second condition in \calle{salg3}
is automatic, but for non-zero central charge, this condition
must be specified independently.  The above algebra implies
$[Q,Z]=[Q^\dagger,Z]=0$, as well as $H\ge |Z|$.

We wish to realize this algebra in a quantum mechanical system.
Our approach is first to present an implementation of this
algebra in a two-sector model, analogous to the
supersymmetric quantum mechanics described in the preceding section, 
and then to
generalize this construction to an arbitrary number of sectors.

To realize the algebra \calle{salg3},
we represent the supercharges as matrices
\beq
\label{Qnew}
Q=\mat{-\eta & 0 \cr A & \eta}\qquad , \qquad
Q^\dagger=\mat{-\eta & A^\dagger \cr 0 & \eta}~~~,
\eeq
where $\eta$ is a real $c$-number.
Then the Hamiltonian and central charge are
determined by the superalgebra to
be, respectively, 
\beqn
\label{Hnew}
H  =  \mat{A^\dagger A+ 2\,\eta^2 & 0\cr 0 & AA^\dagger+2\,\eta^2}
\qquad , \qquad 
Z  =  \mat{2\eta^2 & 0\cr 0 & 2\eta^2}~~.
\eeqn
The operator $\Gamma = \sigma_3$ commutes with the Hamiltonian,
and thus its eigenvalues distinguish the two sectors of the theory.
Notice that the central charge has only non-negative values
in this construction.\footnote{Using complex $\eta$ in fact
generates exactly the same Hamiltonians as using real $\eta$, with
$\eta^2$ replaced by $|\eta|^2$.  The central charge becomes complex,
picking up an overall phase, while the energy bound remains of
the form $H\ge |Z|$.  As a way to refer to this more general setting,
at some points in this paper we use the expression $|Z|$, even though
with our choices, $Z$ has only non-negative values.}

It turns out that the operators that served as supercharges
when there was no central charge, namely
\beq
\tilde Q = \mat{0 & 0 \cr A & 0}~,\qquad 
\tilde Q^\dagger = \mat{0 & A^\dagger \cr 0 & 0}~~,
\eeq
still have a role to play in the centrally extended case.
One notes first that $Q = \tilde Q -\eta\Gamma$, and so 
one can write the Hamiltonian as
\beq
H  = \{\tilde Q,\tilde Q^\dagger\} + |Z|~~~,
\eeq
since $Z=\{\eta\Gamma,\eta\Gamma\}$.  This makes the bound
$H\ge |Z|$ manifest.

When
there is non-vanishing central charge,
$\{Q,\Gamma\}\ne 0$,
and so the supercharges do {\it not} map states from 
one $\Gamma$-sector to the other $\Gamma$-sector. 
The operator $\tilde Q$, on the other hand, not only commutes
with the Hamiltonian and central charge, but also satisfies
$\{\tilde Q, \Gamma\} = 0$.
Therefore, it is the operators $\tilde Q$
and $\tilde Q^\dagger$ that map states from one $\Gamma$-sector
to the other.  Those states for which $H>|Z|$ are doublets under this
operation, while those with $H=|Z|$ are
singlets.

To construct a model with $2N$ sectors for arbitrary
integer $N$, one can
concatenate $N$ two-sector models.  
To construct the supercharges, for example, one places
$2\times 2$ blocks of the form \calle{Qnew} along the diagonal
of a $2N\times 2N$ matrix.  Upon calculating the Hamiltonian
and the central charge, this procedure 
yields a reducible representation of
the centrally extended superalgebra \calle{salg3}.

\begin{figure}
 \begin{center}
 \includegraphics[width=3in]{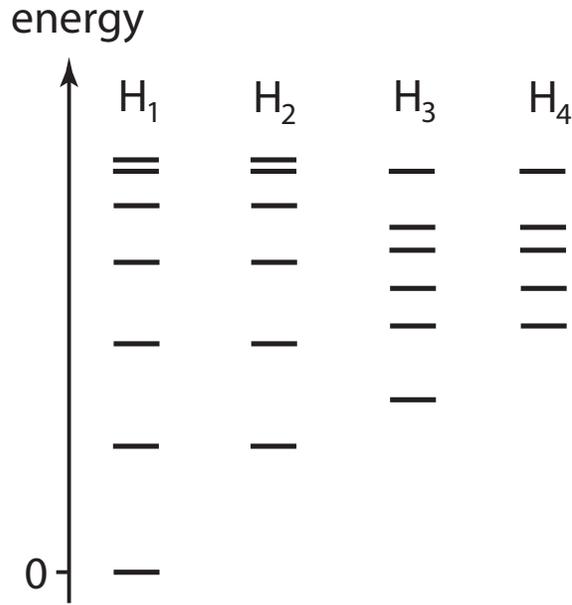}\\[.2in]
 \caption{The energy spectrum of a four-sector model
invariant under supersymmetry with a central charge.
$H_j$ is the Hamiltonian for the $j^{th}$ sector.
Note thtat there is no correlation between the energy levels
of the first partnership (sectors 1 and 2) and those of the
second partnership (sectors 3 and 4).}
 \label{spectrumsplit}
 \end{center}
 \end{figure}

As an example,
a four-sector model has supercharges
\beq
\label{Qfour}
Q = \ba{cc|cc} -\eta_1 & 0 &  & \\  
         A_1 & \eta_1 &  &\\    
\hline
          &  & -\eta_3 & 0  \\
          &  & A_3 & \eta_3  \ea ~~,~~~
Q^\dagger = \ba{cc|cc}-\eta_1 & A_1^\dagger &  & \\
         0 & \eta_1 &  &  \\
\hline
          &  & -\eta_3 & A_3^\dagger  \\
          &  & 0 & \eta_3 \ea~~.
\eeq
The associated Hamiltonian $H$
and central charge $Z$ follow from \calle{salg3}, and are diagonal.
Thus, the
spectrum divides into four sectors, which we number sequentially
along the diagonal.
One notes that sectors 1 and 2 are degenerate, with energies bounded
from below by $2\eta_1^2$, and sectors 3 and 4 are degenerate, with energies
bounded from below by $2\eta_3^2$.  The only exceptions to these degeneracies
are that sectors 1 and 3 each have states that saturate their respective energy bounds, 
while the even sectors do not.  Each of the degenerate pairs of sectors 
we dub a {\it partnership}.  A typical spectrum for a four-sector model is
given in Figure \ref{spectrumsplit}.
 
Once again, it is not the supercharges that swap the degenerate
states within each partnership.  Generalizing from the previous case,
we see that the operators that swap degenerate states within a partnership are
concatenations of the corresponding operators of the two-sector
case; for the four-sector model, these operators 
are
\beq
\label{Qtildefour}
\tilde Q = \ba{cc|cc}0 & 0 &  &    \\
                A_1 & 0 &  &    \\
\hline
                 &  & 0 & 0 \\
                 &  & A_3 & 0 \ea ~~,~~~
\tilde Q^\dagger = \ba{cc|cc}0 & A_1^\dagger &  &   \\
                0 & 0 &  &  \\
\hline
                 &  & 0 & A_3^\dagger   \\
                 &  & 0 & 0 \ea~~.
\eeq

The generalization of this construction to the case of $2N$-sectors exhibits the
essential features we have identified above.  The spectrum divides into
partnerships, consisting of an odd sector and the subsequent even sector.  If $j$ is
an odd integer, then sector $j$ and $j+1$ are bounded from below by
a common constant $\eta_j^2$.  The odd sector has a state with energy
$\eta_j^2$ and the even sector does not, but otherwise these sectors have
degenerate spectra.  One can also
specify generalizations of $\tilde Q$, which are distinct from the supercharges, to
provide the mapping between the degenerate states that reside within each partnership.

Having constructed Hamiltonians invariant under a centrally extended superalgebra,
we now invoke such models in the next section, where they
provide the basis for our analysis of shape
invariance.

\section{The Algebraic Origins of Shape Invariance}

The spectrum of energy levels in supersymmetric quantum mechanics with a central 
charge has, in certain respects, a resemblance to the spectrum of energy levels
that arises in the presence of shape invariance.
For example, comparing Figures \ref{spectrum4}
and \ref{spectrumsplit}, each of which has a spectrum that divides into
four sectors, we notice the following
similarities.  The states
in the first sector (excepting the ground state) are degenerate with those of the
second sector, just as those of third sector (excepting its lowest energy state) are
degenerate with those of the fourth sector.
Furthermore, the lowest energy state of each odd sector satisfies
a first-order equation.  
The obvious generalizations of these statements hold  in the
case of $2N$ sectors.

However, the shape invariant case has two additional features, both
of which
suggest an enhanced algebraic structure.  First, the same
degeneracy pattern that holds within a partnership is present between the 
adjacent even and odd sectors of distinct partnerships.  Second, 
as noted previously, in the shape invariant case, the lowest energy state in 
every sector satisifies a Bogomol'nyi-like first-order equation, something which 
only holds in the odd sectors for the simply supersymmetric case. 

The extra
degeneracies indicate the presence a symmetry operator that not only maps between
degenerate levels within a  partnership, but that also maps between levels
from adjacent sectors that lie in distinct partnerships.  Finding this operator
leads, in turn, to an explanation of the Bogomol'nyi equations.

To approach this problem, we consider first the four-sector case, and then
show that the results so obtained apply to the case of arbitrarily many sectors,
as is necessary if we are to address shape invariance in general.
Using \calle{Qfour}, the four-sector Hamiltonian with centrally extended
supersymmetry can be written as
\beq
\label{Hamfour}
H = \mat{A_1^\dagger A_1 + 2\,\eta_1^2 & 0 & 0 & 0 \cr
                              0 & A_1A_1^\dagger + 2\,\eta_1^2 & 0 & 0\cr
                              0 & 0 & A_3^\dagger A_3+2\,\eta_3^2 & 0 \cr
                              0 & 0 & 0 & A_3A_3^\dagger+2\,\eta_3^2 \cr}~~~.
\eeq
The operator $\tilde Q$ from \calle{Qtildefour} explains the degeneracies
within each partnership that arise
from the superalgebra; this suggests modifying $\tilde Q$
to include an entry that maps the second sector to the third sector, and 
requiring that this 
new
operator be conserved.  
  
We therefore define the ``shift operator'' $S$ by\footnote{The shift operator in shape
invariance should not, of course, be confused with the operation of shapeshifting found
elsewhere
\cite{shapeshifters}.}
\beq
\label{Sfourbyfour}
S\equiv \mat{0 & 0 & 0 & 0  \cr
                A_1 & 0 & 0 & 0  \cr
                0 & C & 0 & 0 \cr
                0 & 0 & A_3 & 0 \cr}~~~,
\eeq
and seek to determine when we can choose $C$ such that $[H,S]=0$.  
As we show below, the shape invariant models correspond to the case
that this is possible  To see this, first we impose the requirement
that $[H,S]=0$, and find that this can be achieved when $A_1^\dagger A_1$ is shape
invariant, and two auxiliary conditions are met, namely that $A_3$ and
$A_1$ are related by a unitary transformation, and that $\eta_3$ is related
to $\eta_1$ so that the energy levels line up suitably.
Appealingly,
the shape invariance condition emerges from the condition that $S$
be conserved.
We then observe that one can look at this result in reverse, concluding that
whenever shape invariance holds for a one-sector Hamiltonian $A_1^\dagger A_1$,
this Hamiltonian can be embedded in a centrally extended supersymmetric
quantum theory with a conserved shift operator, by defining $A_3$ and $\eta_3$
that meet the necessary conditions.

Using the matrix form of $S$ \calle{Sfourbyfour},
the requirement that $H$ and $S$ commute becomes
\beq
\label{Scond}
A^\dagger_3 A_3 C  - C A_1 A_1^\dagger + 2(\eta_3^2-\eta_1^2) C = 0~~~.
\eeq
This condition suggests that there is a simple relation between $A_1$ and
$A_3$.  We therefore suppose that there is a unitary transformation
represented by an operator $\Omega$ such
that
\beq
\label{omegaAA}
A_3 = \Omega^\dagger A_1 \Omega~~~.
\eeq
In order that we are able to make contact with shape invariance,
we allow, and indeed expect, the operator $\Omega$ to implement
a transformation in
parameter space, mapping the $c$-number
parameters of a model (such as $g$ in \calle{H1AA}) to new values,
and thus altering also the values of expressions in the Hamiltonian
(such as $\eta_1^2$ and
$\eta_3^2$) that are functions of these parameters.

The condition that $A_1$ and $A_3$ are unitarily related can be
imposed on
the commutativity condition \calle{Scond}.
Using a unitary operator $U$ such that $U^2=\Omega$, the
resultant equation can be written in the form
\beq
\label{betterScond}
 \Chat \At \Atd - \Atd \At \Chat  = 2 (\bareta_3^2-\bareta_1^2)\Chat~~,
\eeq
where, for simplicity of appearance, 
we have introduced $\At = A_1 U$, $\Chat=UC$, and
$\bareta_j = U\eta_j U^\dagger$.
Our goal is to find a value for $\hatC$ that will lead to a solution
of \calle{betterScond}.  The ansatz with which we have found success
is to choose $\hatC = \At$ (that is, $C = U^\dagger A_1 U$), turning the conservation condition
\calle{betterScond} into
\beq
\label{Scondagain}
\{\,\At\, ,\, [\,\At\, ,\,\Atd\,]\,\} 
= 2(\bareta_3^2-\bareta_1^2)\At~~.
\eeq
Thus, 
to achieve conservation of $S$, the left
side of
\calle{Scondagain} must be proportional to 
$\At$.  We note
that this condition is satisfied if there is a $c$-number
$\kappa$ such that
\beq
\label{comcon}
[\,\At\, ,\,\Atd\,] = \kappa~~~.
\eeq
Combining \calle{comcon} with \calle{Scondagain},
one finds $\{\kappa,\At\}=2(\bareta_3^2-\bareta_1^2)\At$.
Rewriting this in terms of the original quantities,
remembering that $U$ does not commute with $\kappa$,
this relationship takes the form
\beq
\kappa + U^\dagger\kappa U = 2(\eta_3^2-\eta_1^2)~~.
\eeq 
When this condition holds, the requirement
that $S$ commute with $H$ is satisfied.

It is useful re-phrase the above results in reverse.
Suppose that $A_1^\dagger A_1$ is shape invariant.
Then $A_1$ satisfies 
\beq
\label{Shapeatlast}
A_1 A_1^\dagger - U^\dagger A_1^\dagger A_1 U = \kappa~~~,
\eeq
where $\kappa$ is a $c$-number and $U$ implements a shift in the
parameter(s) of the theory, which is precisely the statement \calle{comcon}.
With this condition satisfied, it is possible to construct a multiple sector
model that has $H_1 = A_1^\dagger A_1 + 2 \eta_1^2$ as the Hamiltonian
in its first sector, and that is invariant under
both a centrally extended superalgebra and a shift operator $S$.  
To obtain this mult-sector theory, one defines
quantities $A_3$ and $\eta_3$, respectively, by
\beqn
\label{Shaperesults1}
A_3 & = & (U^\dagger)^2A_1(U)^2\nonumber \\
2\eta_3^2 & = & 2\eta_1^2 + \kappa +U^\dagger \kappa U~~.
\eeqn
The conserved shift operator takes the form
\beq
\label{Sresult}
S=\mat{0 & 0 & 0 & 0\cr A_1 & 0 & 0 & 0 \cr 0 & U^\dagger A_1U & 0 & 0 \cr
       0 & 0 & {U^\dagger}^2 A_1 {U}^2 & 0}~~,
\eeq
while the Hamiltonian for the second sector can be written
as $(U^\dagger A_1 U)^\dagger (U^\dagger A_1 U)+\eta_1^2+\kappa$.
In this way, one sees that the shape invariant theories correspond to 
centrally extended supersymmetric
theories with a conserved shift operator.

It is 
now straightforward to generalize this construction from the four sector
case to a model with an arbitrary 
number of sectors.
Since, due to \calle{Shapeatlast} and \calle{Sresult}, the
relationship between sector $j$ and sector $j+1$ is implemented in
the same way
for each value of $j$ (and not just when the sectors $j$ and $j+1$
fall within a single partnership), the algebraic structure found
above can be readily extended to a theory
with $2N$ sectors, where $N$ is an arbitrary integer.  
For notational compactness, it helps to define the diagonal matrix
$\bub= {\rm diag}(1,U,U^2,U^3,\ldots,U^{2N-1})$ and the
matrix, all of whose entries lie just below the diagonal, 
${\bab}_{i,j} = A_1\delta_{i,j+1}$.  Then in the $2N$ sector model, the 
shift operator (that is, the extra conserved quantity) takes the form
\beq
\label{Sbub}
S = U\bub^\dagger\bab\bub~~~,
\eeq
and it is conserved provided $A_1A_1^\dagger - U^\dagger A_1^\dagger A_1 U =
\kappa$, i.e., provided that $A_1^\dagger A_1$ is shape invariant. Thus, shape invariance
corresponds to the invariance of the multiple-sector Hamiltonian under the 
action of the shift operator $S$.

It is worth noting that conservation of $S$ plays a role here analogous to
that played by the LaPlace-Runge-Lenz vector in the hydrogen atom.  In the hydrogen
atom,
spherical symmetry dictates that the energy eigenvalues depend on a radial
quantum number $n$ and an angular momentum quantum number $\ell$.  The
additional conservation law associated with the LaPlace-Runge-Lenz vector ensures that
states with the same $n$ value but different $\ell$ values are in fact 
degenerate \cite{rungelenz}.
Likewise, when a model is invariant under a centrally extended superalgebra,
this algebra imposes no relation between the energy levels of the different
partnerships; it is conservation of the shift operator that aligns these partnerships
to produce the additional degeneracies that arise in the presence of shape 
invariance.

The example in the appendix shows briefly how the structure we
have derived above applies to a particular case.

\section{Shape Invariance, BPS, and the Shift Operator}

Having obtained the shape invariance condition from the algebra of
centrally extended supersymmetry enhanced by a shift operator, 
we now consider the further implications of this algebra.  For convenience, 
we again consider
initially the four-sector model.   In this case, all states in
the fourth sector are trivially annihilated by $S$.  However, in the
other three sectors,
something more interesting occurs.

Due to \calle{Hamfour}, \calle{Shapeatlast}, and \calle{Shaperesults1}, the 
Hamiltonian of the four-sector model
is related to $S$ in an especially simple way.  In
particular,
\beq
H = S^\dagger S + B~~,
\eeq
where $B$ is a diagonal matrix that, except in its final entry, consists entirely
of $c$-numbers.  One readily determines that
\beq
\label{Bvalues}
B= \mat{2\eta_1^2 & 0 & 0 & 0 \cr
                              0 & 2\eta_1^2+\kappa & 0 & 0\cr
                              0 & 0 & 2\eta_1^2+\kappa + U^\dagger\kappa U & 0 \cr
                              0 & 0 & 0 & H_4 \cr}~~~.
\eeq
In the first three sectors, the energies are constrained
by a Bogomol'nyi bound, $H_k\ge (B)_{kk}$.  This bound is saturated only
for a state
annihilated by $S$; $S$ is a first-order differential operator, and this annihilation
condition then is the Bogomol'nyi equation for these BPS-saturating states.
These states are the ground states of the first three sectors.

If we consider the full four-sector theory,
the identity $S^4=0$ implies that a typical multiplet of
degenerate states
consists of four
states.  The multiplets in which one of the states from the first three
sectors 
satifies
$S\psi=0$ are shortened, however, with one, two, and three states, respectively.
This is analogous to what occurs for
BPS-saturating states when it is the supercharge involved in the annihilation
condition \cite{susyreps} \cite{WittenOlive}.  

Finally, because each of the first three states of the first sector have to be
degenerate with the Bogomol'nyi-saturating ground state of one of the
first three sectors,
the constants in
$B$ represent not only the Bogomol'nyi bounds of the various sectors,
but also
the first three energy eigenvalues of the original Hamiltonian.

Of course, nothing is special about the four-sector model; we can 
easily extend these results to a theory with an arbitrary number
of sectors.  In a model with $2N$ sectors, the BPS structure
still holds, with $S$ defined as in \calle{Sbub}, and
\beqn
\label{HandB}
H & = & S^\dagger S + B \nonumber\\
B & = & {\rm diag}(b_1,b_2,\ldots,b_{2N-1},H_{2n}) \nonumber \\
b_1 & = & 2\eta_1^2 \nonumber \\
b_{j+1} & = & b_j + (U^\dagger)^{j-1}\kappa U^{j-1}~~~. 
\eeqn
In the first $2N-1$ sectors, the ground state saturates a
Bogomol'nyi bound $H=B$ (that is, has energy $b_j$),
and this state is annihilated by the first-order differential
operator $S$.  Because of the
degeneracies produced by conservation of $S$, these Bogomol'nyi 
bound values are also the energies of
the first $2N-1$ states of $H_1$.  These $2N-1$ lowest energy states
of $H_1$ are part of shortened
$S$ multiplets (since $S^{2N}=0$, multiplets of length $2N$ are the
norm); the $j^{th}$ energy level of $H_1$ can be obtained by applying
$S^\dagger$ repeatedly to the
Bogomol'nyi-saturating ground state of the $j^{th}$ sector. While for any finite value of
$2N$, the BPS structure only applies to the first $2N-1$ sectors and energy levels, this
is not a fundamental limitation; as the whole process can be iterated for arbitrarily
large values of $2N$, in fact all the energy levels of $H_1$ (and, indeed, of the
Hamiltonians with which it is associated via shape invariance) fit into this algebraic
framework.

This completes the analysis of the structure of shape invariance.

\section{Summary and Prospects}

We have demonstrated that shape invariance 
is associated with a more
comprehensive invariance algebra: supersymmetry with a
central charge, enhanced by the addition of a shift operator $S$ that maps among
adjacent sectors of the supersymmetric model, even when those sectors come
from distinct partnerships.
Most compellingly, there turns out to be a natural BPS interpretation of 
shape invariance due to this structure.  When the shift operator is conserved,
and hence shape invariance holds, the Hamiltonian can be written
as $H=S^\dagger S + B$,
and so not only is $H\ge B$, but the states for which $H=B$ are the states
annihilated by $S$; the equation $S\psi=0$ is nothing but the Bogomol'nyi
equation for this model.  Finally, in a result that exceeds
conventional BPS results, because $S$ plays a role analogous to a LaPlace-Runge-Lenz
vector, it imposes degeneracies between every pair of adjacent sectors,
and thus the eigenvalues of $B$ are also the
energy eigenvalues of the first sector, and the corresponding states can
be obtained 
by the action of $S^\dagger$ on
the BPS-saturating states.

The algebra we have described gives a natural
framework for understanding the origins of shape invariance.  
Still, it is a curious question as to whether these
Bogomol'nyi bounds can be given a natural topological
explanation \cite{WittenOlive}, with each sector in the shape
invariant case corresponding to a distinct topological sector.  We have pursued 
some initial efforts in this direction, by  using a field theoretic 
approach to study
supersymmetric quantum mechanics with a central charge \cite{FSduality}.  In the context
of the sigma models we have studied in that language, shape invariance amounts to a
restriction on the target space geometry.  Continued efforts should show
if there is additional significance to such a restriction, and whether there
is a natural way to interpret the rest of the algebraic structure described
above in terms of features of the target space.  
We believe such an approach has the potential to lead us to
a topological interpretation of the construction
presented in this paper.

\vspace{.5in}
\leftline{\Large\bf Acknowledgments}
We thank Ted Allen for his generous ear and insightful observations, and
Costas Efthimiou for his helpful comments on the manuscript.

\newpage
\appendix
\section{Appendix} 
\label{appendixA}
As an example of shape invariance, consider the Hamiltonian
\beq
\label{Hsec}
H = -{d^2\over dx^2} + {b\,\sech^2(x)}~~.
\eeq
Setting
\beq
A = {d\over dx}+g\,\tanh(x)~~,
\eeq
one obtains the paired Hamiltonians
\beq
H_1(g) = A^\dagger A = -{d^2\over dx^2} - g(g+1)\,\sech^2(x) + g^2
\eeq
and
\beq
H_2(g) = A A^\dagger = -{d^2\over dx^2} - g(g-1)\,\sech^2(x) + g^2~~.
\eeq
Clearly, $H_2(g)=H_1(g-1)+(2\,g-1)$, which is an explicit
manifestation of the shape invariance condition \calle{ShapeCondition},
recovered in our construction by \calle{Shapeatlast}.

To identify the necessary unitary transformation called for
in our analysis, note that $U^\dagger f(g) U$ must yield $f(g-1)$.  
Such a shift is achieved
by the operator
\beq
U = \exp(\partial/\partial g)~~.
\eeq
The parameter $\kappa$ associated with this model is $\kappa(g) = 2\,g-1$.
The interested reader can easily apply the rest of our construction
to this example.

\bye